\documentclass[amsmath,amssymb,reprint,aps,pra,floatfix]{revtex4-2}
\usepackage{graphicx}
\usepackage{lineno}

\usepackage[utf8]{inputenc}
\usepackage{siunitx}
\DeclareSIUnit{\sqrthertz}{\sqrt{\unit{\hertz}}}
\usepackage[german, english]{babel}

\begin{document}

\title{Hänsch-Couillaud locking of a large Sagnac interferometer: advancing below the flicker floor}

\author{Jannik Zenner}
  \affiliation{Physikalisches Institut, Universität Bonn, Nussallee 12, 53115 Bonn, Germany}
\author{Karl Ulrich Schreiber}
  \affiliation{Research Unit Satellite Geodesy, Technical University of Munich, 80333 Munich, Germany}
  \affiliation{School of Physical Sciences, University of Canterbury, Christchurch 8140, New Zealand}
\author{Simon Stellmer}
  \email{stellmer@uni-bonn.de}
  \affiliation{Physikalisches Institut, Universität Bonn, Nussallee 12, 53115 Bonn, Germany}

\date{\today}

\begin{abstract}

Large Sagnac interferometers in the form of active ring lasers have emerged as unique rotation sensors in the geosciences, where their sensitivity allows to detect geodetic and seismological signals. The passive laser gyroscope variant, however, is still at a stage of development, and thus far, only the Pound-Drever-Hall frequency stabilization technique has been explored, a method limited by residual amplitude modulation. Here, as an alternative method, we present the first Hänsch-Couillaud locked passive laser gyroscope. We find that this method is limited by flicker noise, and we introduce a cost-effective lock-in scheme to overcome this limitation. We achieve a sensitivity of \SI{3.1}{\nano\radian/\second}, corresponding to a fraction of \SI{7.7e-5}{} in the Earth's rotation rate.

\end{abstract}

\maketitle

\section{Introduction}

Various experiments around the world are targeted at highly sensitive rotation measurements based on the Sagnac effect \cite{Sagnac1913}, using meter-sized ring resonators in fields such as geodesy \cite{Schreiber2013, Gebauer2020}, seismology \cite{Igel2005, Igel2011}, and fundamental physics \cite{Stedman1997, Bosi2011}. Most of these sensors \cite{Schreiber2003, DiVirgilio2019, Brotzer2025} are based on the concept of a so-called active ring laser: the laser medium (helium-neon gas) is placed inside the resonator. The frontrunner in terms of sensitivity and stability is the G~ring in Wettzell (Germany) \cite{Schreiber2023, Schreiber2025}.

An alternative concept, the so-called passive laser gyroscope with an external laser feeding two counter-propagating resonator modes, is being explored \cite{Ezekiel1977, Korth2016, Liu2019, Feng2023, Chen2025}. This approach avoids the intra-cavity gain medium and the lasing plasma, both of which deteriorate sensor performance through lock-in and backscatter effects and uncontrolled plasma dynamics.

For a ring-shaped optical resonator with perimeter $P$ enclosing an area $A$ and rotating at a rate $\Omega$, the Sagnac frequency $\delta f$ reads
\begin{align}
    \label{align:sag}
    \delta f = \frac{4 A}{\lambda P} \Omega \sin\theta = \nu_{\text{cw}} - \nu_{\text{ccw}},
\end{align}
where $\lambda$ is the wavelength of the light and $\theta$ is the projection angle of the surface vector onto the rotation vector. Experimentally, for passive laser gyroscopes, the sensitivity of $\delta f$ therefore becomes a question of how well the external laser can be stabilized to the clockwise cavity resonance $\nu_{\text{cw}}$ and the counter-clockwise cavity resonance $\nu_{\text{ccw}}$.

To our knowledge, all passive laser gyroscopes presented thus far employ the Pound-Drever-Hall (PDH) scheme \cite{Black2001} for stabilization of the counter-propagating beams. A key limitation of the PDH method is residual amplitude modulation (RAM), the presence and fluctuation of a non-zero error signal baseline, induced by an asymmetry in phase of the frequency-modulated side bands caused by all kinds of optical components \cite{Kokeyama2014}. A wealth of methods have been presented to reduce the RAM contribution, including active stabilization \cite{Zhang2014}, optical cavity impedance matching \cite{Shi2018}, synthetic triplet generation \cite{Kedar2024}, and correlation analysis \cite{Feng2023b}. Despite these efforts, RAM has not been solved, but at best mitigated. The community is still searching for suitable electronic stabilization methods.

In this work, we examine an alternative to PDH locking in a passive laser gyroscope, namely the Hänsch-Couillaud (HC) locking scheme \cite{hansch1980}. This method can be used for birefringent cavities, like high-finesse ring cavities, utilizing the polarization of the laser beam. Unlike in PDH locking, no frequency-modulation is necessary, thereby completely avoiding the cause of RAM.

\section{Setup}

\begin{figure}[t]
    \includegraphics[width=\columnwidth]{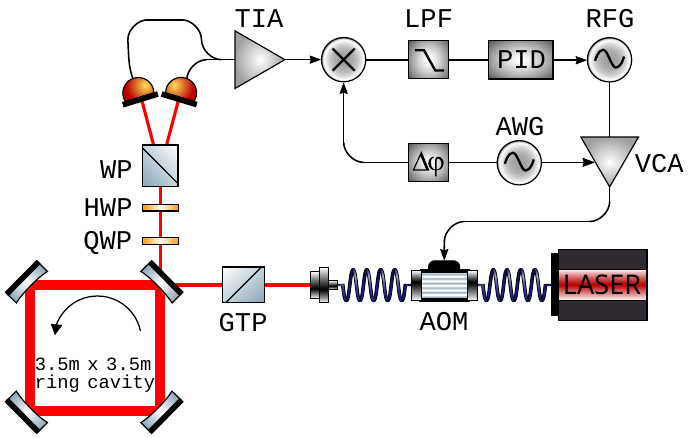}
    \caption{\label{fig:setup} Schematic view of the lock-in Hänsch-Couillaud setup used to stabilize each laser direction onto the ring cavity. Abbreviations: AOM \textit{acousto-optic modulator}, AWG \textit{arbitrary waveform generator}, GTP \textit{Glan-Taylor prism}, HWP \textit{half-wave plate}, LPF \textit{low-pass filter}, PID \textit{proportional-integral-derivative controller}, QWP \textit{quarter-wave plate}, RFG \textit{radio frequency generator}, TIA \textit{transimpedance amplifier}, VCA \textit{voltage controlled attenuator}, WP \textit{Wollaston prism}.}
\end{figure}

The setup is based on a square ring laser cavity of \SI{3.5}{\meter} side length inside a vacuum chamber, described in detail elsewhere \cite{Schreiber2009,Schreiber2006}. The resonator is formed by four mirrors with a radius of curvature of \SI{4}{\meter} and a finesse of about 36,000, as described in Ref.~\cite{Zenner2025}. The free spectral range of the cavity is $f_{\text{FSR}} = c/P =$ \SI{21.4232}{\mega \hertz}, equating to a cavity linewidth $f_{\text{FWHM}} = f_{\text{FSR}}/F$ of about \SI{600}{\hertz}. 

To operate the ring cavity as a passive gyroscope, an external diode laser (TOPTICA DL PRO) at $f_{\text{L}} =$ \SI{473.614}{\tera \hertz} is used. Its light is split into four branches of different intensity levels. 

In one branch, \SI{100}{\micro \watt} of light is frequency-modulated at \SI{5}{\mega \hertz} in an electro-optical modulator, passed through a polarization maintaining fiber, and coupled into the ring cavity. The reflection is detected with a \SI{150}{\mega \hertz} bandwidth photodetector to perform a PDH lock \cite{Black2001} of the diode laser using a fast proportional-integral-derivative (PID) controller (TOPTICA FALC PRO). This pre-stabilization allows to reduce the laser linewidth to about \SI{100}{\hertz} and therefore below $f_{\text{FWHM}}$, and to remove slow frequency drifts originating from  geometrical deformations of the ring cavity. 

Another branch is used to interrogate the laser frequency $f_{\text{L}}$ at a rate of \SI{200}{\hertz} with a wavelength meter. According to this live reading, the ring perimeter is kept constant within a few nanometers using a stacked piezo actuator, as described in Ref.~\cite{Zenner2025}.

The two remaining laser branches are coupled into polarization maintaining optical fibers so that the light can be used to perform the two HC locks, as explained in the following.

\subsection{Hänsch-Couillaud locking method}

The standard HC lock is performed with a simplified version of the setup shown in Fig. \ref{fig:setup} for the counter-clockwise direction and independently in a mirrored way for the clockwise direction at a different corner of the ring cavity.

For each branch, the light is shifted independently by $f_{\text{AOM}} = 9 \cdot f_{\text{FSR}} =$ \SI{192.8088}{\mega \hertz} with a fiber-coupled  acousto-optic modulator (AOM) by applying a radio frequency signal $s_{\text{AOM}}$ at a power level of \SI{30}{\deci \bel m}. This first-order shifted light has an extinction ratio of better than \SI{50}{\deci \bel}, making other deflection orders and the zeroth order negligible for the experiment. The two counter-propagating HC locks are stabilized to the same mode index of the cavity, which is shifted by $9\cdot f_{\text{FSR}}$ with respect to the PDH diode laser lock. The frequencies are generated by direct digital synthesis (DDS) and finetuned via a PID controller connected to the frequency modulation input of the DDS to precisely stabilize the laser frequency to the center of the cavity resonance. The input for the PID controller is produced by the standard HC locking method \cite{hansch1980}.

It is important to note that the principle of HC locking lies in the interrogation of a s-polarized (or any fixed polarization in general) cavity mode with light detuned in polarization. The cavity must therefore be birefringent, which high-finesse ring cavities intrinsically are without the need for polarization optics inside the resonator, as opposed to linear cavities. Using a polarization that is detuned by only a few degrees would yield a better in-coupling efficiency, as all non s-polarized light is always reflected. However, this experiment aims at frequency precision and does not benefit from more light circulation inside the cavity.

The light is linearly polarized at an angle of \SI{45}{\degree} relative to the cavity plane, and the polarization is cleaned at a factor of 100,000:1, where only the extraordinary transmission of a Glan-Taylor prism is used for the experiment. The angle of \SI{45}{\degree} is chosen despite stabilizing the light to a s-polarization mode, as it maximizes the size of the HC error signal and therefore maximizes the signal-to-noise ratio. The p-polarized TEM$_{00}$ mode of the cavity does not interfere, as it is shifted by \SI{4.055}{\mega \hertz} compared to the s-polarized mode, due to a slightly different scale factor induced by the polarization dependent Goos-Hänchen shift \cite{Foster2007, Hurst2017}. The p-polarized mode can easily be distinguished as its finesse is reduced by a factor of more than 100, because the super-polished dielectric mirrors are only highly reflective for s-polarized light.

Each out-coupled beam, at an intensity of \SI{250}{\micro \watt}, is transversely matched to the TEM$_{00}$ mode of the cavity through a single lens and careful alignment of a 5-degrees-of-freedom kinematic mount. The reflected beam passes through a quarter-wave plate and a half-wave plate and is separated at 100,000:1 extinction ratio with a Wollaston prism into the p- and s-polarized parts. All wave plates are true zero-order to minimize polarization drifts induced by temperature variations. The differential detection of the two beams yields the HC error signal. Here, differential detection is achieved by using a cathode-grounded photodiode for one beam and an anode-grounded photodiode for the other, connecting both to the input of the same transimpedance amplifier (gain: \SI{100}{\kilo \volt / \ampere}, bandwidth: \SI{20}{\mega \hertz}). Correctly adjusting the rotation orientation of the wave plates results in an error signal of \SI{\pm 100}{\milli \volt} amplitude symmetrically around zero and with its zero crossing in the center of the linewidth of the cavity mode with frequency $\nu_{\text{cw/ccw}}$. 

This way, locking of both counter-propagating beams is achieved and a Sagnac beat of about \SI{312}{\hertz}, measured by overlapping the transmissions of the ring cavity at a different corner \cite{Zenner2025}, is readily obtained. Investigating the sensitivity of this Sagnac beat however, showed a flicker-noise limitation, as discussed later on.

\subsection{Lock-in method for sub-flicker floor locking}

In an effort to improve the observed flicker noise limitation, the HC lock explained above is enhanced by the lock-in scheme shown in Fig. \ref{fig:setup}. The radio frequency signal $s_{\text{AOM}}$, which has a level of \SI{30}{\deci \bel m} at the frequency $f_{\text{AOM}}$, is amplitude-modulated with the modulation frequency $f_{\text{m}} =$ \SI{10}{\kilo \hertz}:
\begin{align}
    \label{align:modulation}
    s_{\text{AOM}}(t) = \frac{1}{2}[1+ m\cdot \cos(2\pi f_{\text{m}}t)]\cdot s_{\text{AOM}},
\end{align}
where the modulation index $m$ is very close to 1. A voltage-controlled RF attenuator is used for the modulation. The sinusoidal signal of \SI{10}{\volt} amplitude is created by an arbitrary waveform generator. The attenuation is between \SI{60}{\deci \bel} at \SI{0}{\volt} and \SI{4}{\deci \bel} at \SI{10}{\volt} onto the RFG signal of \SI{34}{\deci \bel m}. The chosen modulation frequency $f_{\text{m}}$ equals a period time of \SI{100}{\micro \second}, which is shorter than the cavity ringdown time of $\tau_{dec}=$\SI{270}{\micro \second}. The intra-cavity power and the transmission used for the Sagnac beat are only modulated by \SI{5.9}{\percent}, as the cavity acts as a low-pass filter with a damping of the form $1/\tau_{dec} \cdot e^{-t/\tau_{dec}}$.

The amplitude-modulated beam is reflected off the cavity mirror, transmits the wave plates, is separated in polarization in the Wollaston prism and is detected with the same photodiodes and transimpedance amplifier as in the standard method explained above. To retrieve the error signal, demodulation is done by mixing the photodiode signal with a correctly phase-shifted modulation frequency $f_{\text{m}}$. The phase shift $\Delta \varphi$ is manually set to be in phase with the detected signal's amplitude modulation within \SI{1}{\degree} in order to maximize the amplitude of the demodulated error signal. Residual modulation is rejected by a Butterworth low-pass filter at a \SI{3}{\deci \bel} threshold frequency of \SI{2.2}{\kilo \hertz}.

\section{Results}

\begin{figure} [t]
    \includegraphics[width=\columnwidth]{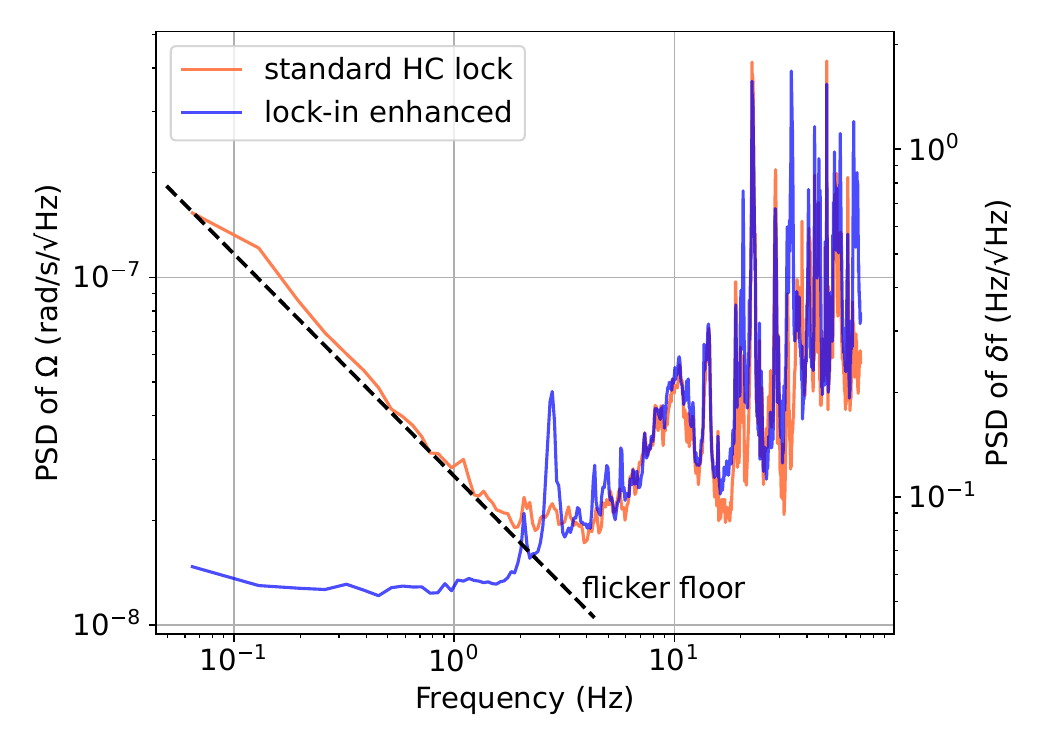}
    \caption{\label{fig:psd} Power spectral densities of the measured Sagnac frequency. The standard HC lock (orange) clearly is flicker noise dominated in the low frequency regime up until about \SI{2}{\Hz}. For the lock-in method (blue), noise contributions in this regime are purely shot-noise like.}
\end{figure}

\begin{figure} [t]
    \includegraphics[width=\columnwidth]{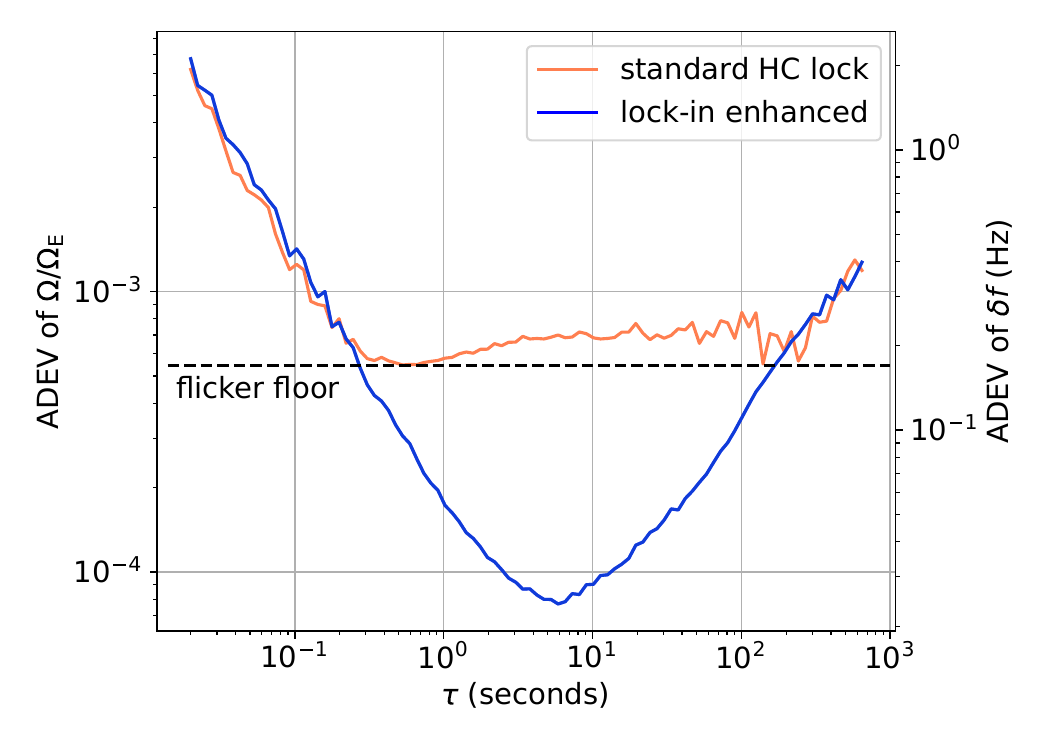}
    \caption{\label{fig:adev} Classic Allan deviations of the measured Sagnac frequency. The lock-in method (blue) beats the standard HC lock (orange) at averaging times between \SI{0.3}{\second} and \SI{200}{\second} as it is not flicker noise limited. It reaches a minimum at about \SI{6}{\second} of \SI{7.7e-5}{}$\,\Omega_{\text{E}}$ or \SI{24}{\milli \hertz}.}
\end{figure}

To assess the performance of a HC locked passive laser gyroscope, two half-hour Sagnac measurements are performed in direct succession, one with the standard HC locking method and one with the lock-in enhanced HC method.

To analyze noise contributions, the Welch power spectral density (PSD) \cite{Welch1967} is computed for both datasets and shown in Fig.~\ref{fig:psd}. It can clearly be seen that the predominant kind of noise for the normal HC method in the regime below about \SI{2}{\hertz} is of the $1/f$-type (flicker noise). Noise of this type has long been observed and studied in DC-type detection schemes in photodiodes \cite{Rubiola2006}, originating mainly from dark currents induced by defects in the semiconductor material \cite{Kinch2005}. Additionally, further flicker-type noise contributions like laser power modulations cannot be excluded. When enhancing the locking scheme with a lock-in detection of the error signal, the detection is shifted to the modulation frequency and the flicker noise is suppressed, leading to a purely shot-noise limited PSD at a level of about \SI{1.2e-8}{\radian/\second/\sqrthertz} for the rotation rate measurement. At higher frequencies, both methods yield the same noise characteristics. We observe three dominant noise peaks. The peaks at \SI{23}{\hertz} and \SI{34}{\hertz} are of acoustic origin and attributed to cooling fans in the lab, as verified by accelerometer measurements. The third peak at \SI{50}{\hertz} is caused by electronics. 

Furthermore, the classic Allan deviation \cite{Allan1966} is calculated to examine the rotation rate measurement performance, as shown in Fig.~\ref{fig:adev}. When lock-in modulating, the best sensitivity is reached at about \SI{6}{\second} with \SI{7.7e-5}{}$\,\Omega_{\text{E}}$ or \SI{3.1}{\nano\radian/\second}. This is similar as previously reported for a similar sized passive laser gyroscope using the PDH locking technique \cite{Feng2023}. This equates to a deviation minimum of the Sagnac frequency $\delta f$ of \SI{24}{\milli \hertz}. Each of the two symmetrically assembled counter-propagating locks is therefore stabilized to a level of \SI{17}{\milli \hertz}, which corresponds to a fraction of \SI{2.8e-5}{} in the cavity linewidth $f_{\text{FWHM}}$ and a fraction of \SI{3.6e-17}{} of the absolute laser frequency $f_{\text{L}}$.

The gyroscope using the standard HC lock does not reach this level of sensitivity. The flicker floor at \SI{5.5e-4}{}$\,\Omega_{\text{E}}$, corresponding to a \SI{170}{\milli\hertz} deviation of $\delta f$, limits the Allan deviation from reaching lower values for integration times $\tau$ between \SI{0.2}{\second} and \SI{200}{\second}. Independent of which of the two locking types is used, the drift for integration times larger than \SI{200}{\second} is at an equal random walk level. This drift is assumed to be caused by the optics and optomechanics used in the HC setup: Temperature and pressure drifts distort the geometrical shape of the quarter- and half-wave plates at fractions of the wavelength, resulting in a varying influence on the beam's polarization. This results in a shift in the intensities of the beams separated by the Wollaston prism and therefore a baseline drift in the generated error signal.

\section{Conclusion}

We have presented the first Hänsch-Couillaud locked passive laser gyroscope to explore an alternative to the already established PDH locking scheme. The experiment shows that this method is intrinsically limited by flicker noise when generating the error signal. With a simple lock-in setup demonstrated here, this flicker floor can be overcome to produce a rotation rate measurement at the \SI{7.7e-5}{}\,$\Omega_{\text{E}}$ level, comparable to a PDH locked large passive gyroscope of similar size. As this lock-in enhancement is simple and cost-effective to implement, it might find use in other fields where the HC scheme is being used. The main limitation is drift on time scales longer than 10\,s, likely driven by temperature fluctuations. Better control over environmental parameters will improve the sensitivity substantially.

\section*{Acknowledgments}
We are indebted to U.~Hugentobler and the Forschungseinrichtung Satellitengeodäsie at TU Munich for the loan of the ring laser hardware. We acknowledge fruitful discussions with T.~Koch, Th.~Gereons, J.~Kodet, H.~Igel, and A.~Brotzer, as well as experimental support from P.~Hänisch. \\

\section*{Data Availability Statement}
The data that support the findings of this study are available from the corresponding author upon reasonable request.

\section*{Disclosures}
The authors declare no conflicts of interest.

\section*{Funding}
We acknowledge financial support from the European Research Council ERC through grants No.~101123334 and 101213032.

\bibliography{bib}

\end{document}